REMARKS:
Below you find the *LaTeX source* of our recent paper.
The accompanying figures (in PostScript) are distributed separately in
an uuencoded compressed tar-file as required by the rules for
PostScript files on this bulletin board.

The LaTeX distribution is wrapped up in a single file with "--- snip ---"
lines delimiting the individual files. The corresponding file names are
indicated in the line ">>> <filename <<<" just above the leading "---
snip ---" lines.

The *full* distribution consists of the following files:

LaTeX part (this file):
-----------------------
2ladmewndr.tex:
   LaTeX source of the paper.

artcustom.sty, a4wide.sty, a4.sty:
   Required style files except epsf.sty. epsf.sty is of any use only in
   connection with the Rokicki dvips converter. a4wide.sty uses this
   particular version of a4.sty (with the later separately available at
   this bulletin board although I don't know whether the two are compatible).

PostScript part (distributed separately):
-----------------------------------------
fig1ew.ps, fig2ew.ps, fig3ew.ps, fig4ew.ps, fig5ew.ps:
   PostScript files read in by "2ladmewndr.tex" via epsf.sty provided with
   the Rokicki dvips DVI-to-PostScript converter. If you don't have the
   Rokicki dvips you should comment out the relevant epsf-commands in
   "2ladmewndr.tex".

Best regards,
             MARKUS LAUTENBACHER

MARKUS E. LAUTENBACHER,         | phone: +49/89/3209-2387
Technical University Munich,    | fax:   +49/89/32092296
Physics Dept.-Theo. Physics T31,| Internet:
D-8046 Garching, FRG.           | lauten@feynman.t30.physik.tu-muenchen.de

                           >>> 2ladmewndr.tex <<<
--- snip --- snip --- snip --- snip --- snip --- snip --- snip --- snip ---

\documentstyle[12pt,a4wide,artcustom,epsf]{article}

\newcommand{\dS}{\Delta S=1}
\newcommand{\dB}{\Delta B=1}

\newcommand{\ord}{{\cal O}}

\newcommand{\as}{\alpha_{\rm s}}
\newcommand{\aem}{\alpha}

\newcommand{\gf}{\gamma_5}

\newcommand{\hg}{\hat{\gamma}}

\newcommand{\gs}{\hat{\gamma}_{\rm s}^{(0)}}
\newcommand{\gem}{\hat{\gamma}_{\rm e}^{(0)}}

\newcommand{\gss}{\hat{\gamma}_{\rm s}^{(1)}}
\newcommand{\gse}{\hat{\gamma}_{\rm se}^{(1)}}

\newcommand{\gssndr}{\hat{\gamma}_{\rm s,NDR}^{(1)}}
\newcommand{\gsshv }{\hat{\gamma}_{\rm s,HV }^{(1)}}
\newcommand{\gsendr}{\hat{\gamma}_{\rm se,NDR}^{(1)}}
\newcommand{\gsehv }{\hat{\gamma}_{\rm se,HV }^{(1)}}

\newcommand{\vQ}{\vec{Q}}

\newcommand{\hZ}{\hat{Z}}

\newcommand{\eps}{\varepsilon}

\newcommand{\brackcc}[1]{\left[#1\right]_{\rm cc}}
\newcommand{\brackp }[1]{\left[#1\right]_{\rm p}}
\newcommand{\brackets}[2]{\left[ #1 \right]_{\rm #2}}

\newcommand{\re}{\hat{r}_{\rm e}}
\newcommand{\rs}{\hat{r}_{\rm s}}
\newcommand{\dre}{\Delta\hat{r}_{\rm e}}
\newcommand{\drs}{\Delta\hat{r}_{\rm s}}

\newcommand{\svs}{\vbox{\vskip 5mm}}

\newcommand{\msvs}{\vbox{\vskip 7mm}}

\newcommand{\nn}{\nonumber}
\newcommand{\eqn}[1]{(\ref{#1})}

\newcommand{\newsection}[1]{\section{#1}\setcounter{equation}{0}}

\newcommand{\doctype}{}

\newcommand{\PreprintOrPaper}[2]{ \ifnum \doctype=1 #1 \else #2 \fi }

\begin{document}


\paperid{
{\sf MPI-PAE/PTh 107/92}\\
{\sf TUM-T31-30/92}
}

\date{\sf November 1992}

\author{\\
{\normalsize Andrzej J. BURAS${}^{1,2}$, Matthias JAMIN${}^{3}$
and Markus E. LAUTENBACHER${}^{1}$}\\
\ \\
{\small\sl ${}^{1}$ Physik Department, Technische Universit\"at
M\"unchen,
                    D-8046 Garching, FRG.}\\
{\small\sl ${}^{2}$ Max-Planck-Institut f\"ur Physik
                    -- Werner-Heisenberg-Institut,}\\
{\small\sl P.O. Box 40 12 12, D-8000 M\"unchen, FRG.}\\
{\small\sl ${}^{3}$ Division TH, CERN, 1211 Geneva 23, Switzerland.}}

\title{
{\LARGE\sf
Two--Loop Anomalous Dimension Matrix for\\
$\dS$ Weak Non-Leptonic Decays {\rm II}:
$\ord(\aem\,\as)$}\footnote{Supported by the German
Bundesministerium f\"ur Forschung und Technologie under contract 06 TM
732 and by the CEC Science project SC1-CT91-0729.}
}

\maketitle

\begin{abstract}
\noindent
We calculate the $10\times 10$ two--loop anomalous dimension matrix to
order $\ord(\aem\,\as)$ in the dimensional regularization scheme with
anticommuting $\gf$ (NDR) which is necessary for the extension of the
$\dS$ weak Hamiltonian involving electroweak penguins beyond the
leading logarithmic approximation. We demonstrate, how a direct
calculation of penguin diagrams involving $\gf$ in closed fermion loops
can be avoided thus allowing a consistent calculation of two--loop
anomalous dimensions in the simplest renormalization scheme with
anticommuting $\gf$ in $D$ dimensions.
We give the necessary one--loop finite terms which allow to obtain the
corresponding two--loop anomalous dimension matrix in the HV scheme with
non--anticommuting $\gf$.
\end{abstract}

\newpage
\setcounter{page}{1}


\newsection{Introduction}
The inclusion of next--to--leading QCD corrections to the effective low
energy Hamiltonian for non--leptonic decays requires the calculation of
the relevant two--loop anomalous dimension matrices. In the case of the
$\dS$ Hamiltonian there are ten operators $Q_i, i=1,\ldots,10$ to be
considered: current--current operators $(i=1,2)$, QCD penguin operators
$(i=3,\ldots,6)$ and electroweak penguin operators $(i=7,\ldots,10)$.
Consequently, one deals with $10\times10$ anomalous dimension matrices.

Because of the presence of electroweak penguin operators with Wilson
coefficients of $\ord(\aem)$, a consistent analysis must involve anomalous
dimensions resulting from both strong and electromagnetic interactions.
Working to first order in $\aem$ but to all orders in $\as$, the following
anomalous dimension matrix is needed for the leading and next--to--leading
logarithmic approximation for the Wilson coefficient functions,
\begin{equation}
\hg \; = \;
\frac{\as}{4\pi}\, \gs +
\frac{\aem}{4\pi}\, \gem +
\frac{\as^2}{(4\pi)^2}\, \gss +
\frac{\aem\,\as}{(4\pi)^2}\, \gse \,.
\label{eq:1.1}
\end{equation}

The one--loop matrices $\gs$ and $\gem$ have been calculated long time
ago
\cite{gaillard:74}--\nocite{altarelli:74,vainshtein:77,gilman:79,guberina:80,bijnenswise:84,burasgerard:87,sharpe:87,lusignoli:89,flynn:89}\cite{buchallaetal:90}.
The $2\times2$ submatrix of the two--loop QCD matrix $\gss$ involving
current--current operators $Q_1$ and $Q_2$ has been calculated in
ref.~\cite{altarelli:81,burasweisz:90}. Recently, we have generalized
these two--loop calculations to the penguin operators $Q_i,
i=3,\ldots,10$ \cite{burasetal:92a,burasetal:92b}, so that the
full matrix $\gss$ is also known. The purpose of the present paper is
the calculation of $\gse$.

The calculation of $\gse$ proceeds in analogy to $\gss$. In fact all the
singularities calculated in ref.~\cite{burasetal:92b} can be used here
so that our main task was the calculation of the relevant colour and
electric charge factors in two--loop diagrams involving one gluon and
one photon. Yet as will be seen the structure of basic expressions and
of the results differs from the one found in the pure QCD case, because
now the electric charges of quarks matter and the flavour symmetries
present in ref.~\cite{burasetal:92b} are broken. In fact this breakdown
of flavour symmetry is the origin of the existence of electroweak
penguin operators.

In order to make the comparison with the calculation of $\gss$ as easy
as possible, we have organized the present paper in a similar way as it
was done in ref.~\cite{burasetal:92b}. We present here however only the
results in the naive dimensional regularization with anticommuting $\gf$
(NDR). The consistency of the NDR and 't Hooft--Veltman (HV) scheme for
the calculation at hand has been demonstrated in
ref.~\cite{burasetal:92b}. There we have developed a method which
allows to avoid a direct calculation of penguin diagrams involving
$\gf$ in closed fermion loops which are ambiguous in the case of the
NDR scheme. Our method can be generalized to the present case so that
an unambiguous calculation of $\gse$ can be accomplished in the NDR
scheme. This is gratifying because this scheme is certainly the most
convenient and contrary to the HV scheme
\cite{trueman:79,burasweisz:90,barrosoetal:92} satisfies all
Ward--identities in the context of the minimal subtraction scheme.

Our paper is organized as follows: In
section 2, we recall the explicit expressions for the operators $Q_i$,
and we classify the one-- and two--loop diagrams into current--current
and penguin diagrams. We also discuss the basic formalism necessary for
the calculation of $\gse$. In section 3, we recall the matrix $\gem$.
In section 4, the calculations and results for two--loop
current--current diagrams are presented. In section 5, an analogous
presentation is given for two--loop penguin diagrams. In section 6, we
combine the results of the previous sections to obtain $\gse$ in the
NDR scheme.  We discuss various properties of this matrix, in
particular its Large--N limit.  Section 7 contains a brief summary of
our paper.  In appendices A and B, explicit expressions for the elements
of the $10 \times 10$ matrices $\gem$ and $\gse$ for arbitrary numbers
of colours ($N$) and flavours ($f$) are given. In appendix~C, the
corresponding results for $\gse$ in the case of $N=3$ are presented.

\newsection{General Formalism}
\subsection{Operators}

The ten operators considered in this paper are given as follows
\begin{eqnarray}
Q_{1} & = & \left( \bar s_{\alpha} u_{\beta}  \right)_{\rm V-A}
            \left( \bar u_{\beta}  d_{\alpha} \right)_{\rm V-A}
\, , \nn \\
Q_{2} & = & \left( \bar s u \right)_{\rm V-A}
            \left( \bar u d \right)_{\rm V-A}
\, , \nn \\
Q_{3} & = & \left( \bar s d \right)_{\rm V-A}
   \sum_{q} \left( \bar q q \right)_{\rm V-A}
\, , \nn \\
Q_{4} & = & \left( \bar s_{\alpha} d_{\beta}  \right)_{\rm V-A}
   \sum_{q} \left( \bar q_{\beta}  q_{\alpha} \right)_{\rm V-A}
\, , \nn \\
Q_{5} & = & \left( \bar s d \right)_{\rm V-A}
   \sum_{q} \left( \bar q q \right)_{\rm V+A}
\, , \nn \\
Q_{6} & = & \left( \bar s_{\alpha} d_{\beta}  \right)_{\rm V-A}
   \sum_{q} \left( \bar q_{\beta}  q_{\alpha} \right)_{\rm V+A}
\, , \label{eq:2.1} \\
Q_{7} & = & \frac{3}{2} \left( \bar s d \right)_{\rm V-A}
         \sum_{q} e_{q} \left( \bar q q \right)_{\rm V+A}
\, , \nn \\
Q_{8} & = & \frac{3}{2} \left( \bar s_{\alpha} d_{\beta} \right)_{\rm V-A}
         \sum_{q} e_{q} \left( \bar q_{\beta}  q_{\alpha}\right)_{\rm V+A}
\, , \nn \\
Q_{9} & = & \frac{3}{2} \left( \bar s d \right)_{\rm V-A}
         \sum_{q} e_{q} \left( \bar q q \right)_{\rm V-A}
\, , \nn \\
Q_{10}& = & \frac{3}{2} \left( \bar s_{\alpha} d_{\beta} \right)_{\rm V-A}
         \sum_{q} e_{q} \left( \bar q_{\beta}  q_{\alpha}\right)_{\rm V-A}
\, , \nn
\end{eqnarray}
where $\alpha$, $\beta$ denote colour indices ($\alpha,\beta
=1,\ldots,N$) and $e_{q}$  are quark charges. We omit the colour indices
for the colour singlet operators. $(V\pm A)$ refer to $\gamma_{\mu} (1
\pm \gf)$. This basis closes  under QCD and QED renormalization.

At one stage it
will be useful to study other bases, in particular the basis in which
the first two operators are replaced by their Fierz conjugates,
\begin{eqnarray}
\widetilde{Q}_{1} & = &
\left( \bar s d \right)_{\rm V-A}
\left( \bar u u \right)_{\rm V-A}
\, , \nn \\
\widetilde{Q}_{2} & = &
\left( \bar s_{\alpha} d_{\beta}  \right)_{\rm V-A}
\left( \bar u_{\beta}  u_{\alpha} \right)_{\rm V-A}
\, ,
\label{eq:2.2}
\end{eqnarray}
with remaining operators unchanged. In fact the latter basis is the one
used by Gilman and Wise \cite{gilman:79}. We prefer however to put
$Q_{2}$ in the colour singlet form as in eq.~\eqn{eq:2.1}, because it
is this form in which this operator enters the tree level Hamiltonian.
Let us finally recall that the Fierz conjugates of the $(V-A)\otimes
(V+A)$ operators $Q_{i}, \; i=5,\ldots,8$ are given by
\begin{eqnarray}
\widetilde{Q}_{6} & = & -\,8
\sum_{q} \left( \bar s_{\rm L} q_{\rm R} \right)
         \left( \bar q_{\rm R} d_{\rm L} \right)
\, , \nn \\
\widetilde{Q}_{8} & = & -\,12
\sum_{q} e_{q} \left( \bar s_{\rm L} q_{\rm R} \right)
               \left( \bar q_{\rm R} d_{\rm L} \right)
\, ,
\label{eq:2.3}
\end{eqnarray}
with similar expressions for $\widetilde{Q}_{5}$ and $\widetilde{Q}_{7}$.
Here $q_{R,L} = \frac{1}{2} (1\pm\gf) q$. The Fierz conjugates of the
$(V-A)\otimes (V-A)$ penguin operators $Q_{3}, Q_{4}, Q_{9}$, and
$Q_{10}$  to be denoted by $\widetilde{Q}_{i}, \; i=3,4,9,10$ are found in
analogy to \eqn{eq:2.2}.

\subsection{Classification of Diagrams}

In order to calculate the anomalous dimension matrices $\gem$ and
$\gse$, one has to insert the operators of eq.~\eqn{eq:2.1} in
appropriate four--point functions and extract $1/\eps$ divergences.
The precise relation between $1/\eps$  divergences in one-- and
two--loop diagrams and one-- and two--loop anomalous dimension matrices
will be given in the following subsection. Here, let us only recall that
insertion of any of the operators of eq.~\eqn{eq:2.1} into the diagrams
discussed below results into a linear combination of the operators $Q_{i}$.
The row in the anomalous dimension matrix corresponding to the inserted
operator can then be obtained from the coefficients in the linear
combination in question.

%

There are three basic ways a given operator can be inserted in a
four--point function. They are shown in fig.~\ref{fig:1}, where the dot
denotes the interaction described by a current in a given
operator of eq.~\eqn{eq:2.1} and the wavy line denotes a gluon or a
photon.

We will refer to the insertions of fig.~\ref{fig:1}(a) as
``current--current'' insertions. The insertions of fig.~\ref{fig:1}(b)
and (c) will then be called ``penguin insertions'' of type 1
and type 2 respectively.

The complete list of diagrams necessary for one-- and two--loop
calculations is given in figs.~\ref{fig:2}--\ref{fig:5}. At one--loop
level one has three current--current diagrams (fig.~\ref{fig:2}) to be
denoted as in ref. \cite{burasweisz:90} by $D_{1}$ -- $D_{3}$, and one
penguin diagram (fig.~\ref{fig:3}) of each type to be denoted by
$P_{0}^{(1)}$ and $P_{0}^{(2)}$ for type 1 and type 2 insertions
respectively. At two--loop level there are 21 current--current diagrams
shown in fig.~\ref{fig:4}, to be denoted by $D_{4}$ -- $D_{24}$ , and 12
penguin diagrams of each type shown in fig.~\ref{fig:5} to be denoted
by $P_{1}^{(1)}$ --$P_{14}^{(1)}$ and $P_{1}^{(2)}$ --$P_{14}^{(2)}$
respectively. The notation is similar to ref.\cite{burasetal:92b}, but
this time the wavy lines denote a gluon or photon so that complete
$\ord(\aem\,\as)$ is obtained. The omitted diagrams $P_6$ and $P_7$
having triple boson vertices do not contribute here.
The penguin diagrams which have no $1/\eps$ divergences and some
examples of penguin diagrams which vanish identically in dimensional
regularization are given in figs.~6 and 7 of ref.~\cite{burasetal:92b}.
It is needless to say that all possible permutations of gluons and
photons have to be considered as well as left--right reflections. In
the case of current--currrent diagrams also up--down reflections have
to be considered.

\subsection{Basic Formulae for Anomalous Dimensions}

The anomalous dimensions of the operators $Q_i$, calculated in the
$\overline{\rm MS}$ scheme, are obtained from the $1/\eps$
divergences of the relevant one-- and two--loop diagrams with $Q_i$
insertions and from the $1/\eps$ divergences in the quark
wave--function renormalization. Let us denote by $\vQ$ a column
vector composed of operators $Q_i$. Then
\begin{equation}
\hat{\gamma}(g,e) =
\hZ^{-1} \mu \frac{\partial}{\partial \mu} \hZ,
\qquad
\vQ_{\rm B} = \hZ \; \vQ \,,
\label{eq:2.4}
\end{equation}
where $\vQ_{\rm B}$ stands for bare operators. Working in
$D=4-2\,\eps$ dimensions, we can expand $\hZ$ in inverse powers of $\eps$
as follows
\begin{equation}
\hZ = \hat{1} + \sum_{k=1}^{\infty} \frac{1}{\eps^k} \hZ_k(g,e)\,,
\label{eq:2.5}
\end{equation}
where $g$ and $e$ are the QCD and QED renormalized coupling constants.

Inserting \eqn{eq:2.5} into \eqn{eq:2.4}, one derives a useful result
\begin{equation}
\hat{\gamma}(g,e) \; = \;
-\,2 g^2 \frac{\partial \hZ_1(g,e)}{\partial g^2}
-2 e^2 \frac{\partial \hZ_1(g,e)}{\partial e^2} \,.
\label{eq:2.6}
\end{equation}

Let us next denote by $\Gamma^{(4)}(\vec{Q})$ and
$\Gamma_{B}^{(4)}(\vec{Q}^{B})$ the renormalized and the bare
four--quark Green functions with operator $\vec{Q}$ insertions.
Strictly speaking $\Gamma^{(4)}(\vec{Q})$ and
$\Gamma_{B}^{(4)}(\vec{Q}^{B})$ are matrices, because the insertion of
a single operator in a given diagram results in a linear combination of
operators.

At the one--loop level $\Gamma_{B}^{(4)}$ is obtained by evaluating the
diagrams of figs.~2 and 3. At the two--loop level it is found by
evaluating the diagrams of figs.~4 and 5 and subtracting the
corresponding two--loop counter terms. Next
\begin{equation}
\Gamma^{(4)}(\vQ) = \hZ_\psi \hZ^{-1} \Gamma_{\rm B}^{(4)}(\vQ_{\rm B})\,,
\label{eq:2.7}
\end{equation}
where $\hZ_\psi$ is a matrix which represents the renormalization of the
four quark fields on external lines. In the pure QCD case this matrix
was diagonal. The $\ord(\aem)$ and $\ord(\aem\,\as)$ terms in this matrix
are however non--diagonal, because in the case of penguin operators
the $d \bar d$, $s \bar s$, and $b \bar b$ parts are renormalized
differently from $u \bar u$ and $c \bar c$ parts. We next expand
$\Gamma_{\rm B}^{(4)}$ and $\hZ_\psi$ in inverse powers of $\eps$
as follows
\begin{equation}
\hZ_\psi =
\hat{1} + \sum_{k=1}^{\infty} \frac{1}{\eps^k} \hZ_{\psi,k}(g,e)\,,
\label{eq:2.8}
\end{equation}
\begin{equation}
\Gamma_{\rm B}^{(4)}(Q_i) =
1 + \sum_{k=1}^{\infty} \frac{1}{\eps^k} Z^{(\Gamma)}_{k}(g,e,Q_i)
+ {\rm finite} \,,
\label{eq:2.9}
\end{equation}
where
\begin{equation}
\hZ_{\psi,1}(g,e) =
\frac{e^2}{(4\pi)^2} \hZ_\psi^{(0)} +
\frac{e^2\, g^2}{(4\pi)^4} \hZ_\psi^{(1)} + \ldots\,,
\label{eq:2.10}
\end{equation}
\begin{equation}
Z^{(\Gamma)}_{1}(g,e,Q_i) = \sum_{j=1}^{10} \left(
\frac{e^2}{(4\pi)^2} (d_1)_{ij} +
\frac{e^2\, g^2}{(4\pi)^4} (d_2)_{ij} + \ldots
\right)\,,
\label{eq:2.11}
\end{equation}
where the dots stand for the pure QCD case which has been already
considered in ref.~\cite{burasetal:92b} and higher order corrections.

Demanding $\Gamma^{(4)}$ to be finite, we find $\hZ_1(g,e)$ and
using \eqn{eq:2.6}
\begin{eqnarray}
\left(\gem\right)_{ij} & = &
 -\,2 \left[ (\hZ_\psi^{(0)})_{ij} + (d_1)_{ij} \right] \,,
\label{eq:2.12} \\
\left(\gse\right)_{ij} & = &
 -\,4 \left[ (\hZ_\psi^{(1)})_{ij} + (d_2)_{ij} \right] \,.
\label{eq:2.13}
\end{eqnarray}
The matrices $\hZ_\psi^{(0)}$ and $\hZ_\psi^{(1)}$ are given in
section 2.5.

\subsection{Renormalization Scheme Dependence of $\gse$}

The two--loop anomalous dimension matrix $\gse$ depends on the
renormalization scheme for operators and in particular on the treatment
of $\gf$ in $D\not= 4$ dimensions. This is signaled by the scheme
dependence of the finite terms $\rs$ and $\re$ in the renormalized
one--loop matrix elements
\begin{equation}
\langle \vQ \rangle =
\left[ \hat{1} + \frac{\as}{4\pi} \rs + \frac{\aem}{4\pi} \re \right]
\langle \vec{Q}^{(0)} \rangle \,,
\label{eq:2.14}
\end{equation}
where $\langle \vec{Q}^{(0)} \rangle$ denotes tree--level matrix
elements.

If $\hZ_a$ and $\hZ_b$ are two renormalization factors of \eqn{eq:2.4}
corresponding to two renormalization schemes $a$ and $b$, then
\begin{equation}
\hZ_a = \hZ_b
\left[ \hat{1} + \frac{\as}{4\pi} \drs + \frac{\aem}{4\pi} \dre \right] \,,
\label{eq:2.15}
\end{equation}
where
\begin{equation}
\drs = (\rs)_b - (\rs)_a \, ,
\qquad
\dre = (\re)_b - (\re)_a \, .
\label{eq:2.16}
\end{equation}

Using eqs.~\eqn{eq:2.15} and \eqn{eq:2.4} one finds the relation between
$\gse$ calculated in schemes $a$ and $b$.
\begin{equation}
\left(\gse\right)_b \; = \; \left(\gse\right)_a +
\left[ \drs, \gem \right] + \left[ \dre, \gs \right] \, .
\label{eq:2.17}
\end{equation}

This relation is very useful as it allows to test the compatibility of the
two--loop anomalous dimensions calculated in different renormalization
schemes and plays a role in the proof of scheme independence of
physical quantities as demonstrated in ref.~\cite{burasetal:92d}.

\subsection{The Matrices $\hZ_\psi^{(0)}$ and $\hZ_\psi^{(1)}$}

We give here the matrices $\hZ_\psi^{(0)}$ and $\hZ_\psi^{(1)}$ in the
Feynman gauge. The wave--function renormalization for a quark of charge
$q$ is given in the Feynman gauge by
\begin{equation}
Z^{(\psi)} =
1 - \frac{\aem}{4\pi} \frac{1}{\eps}\, q^2
+ \frac{\aem\,\as}{(4\pi)^2} \frac{1}{\eps}\,\frac{3}{2}\, q^2 C_F+\ldots \,,
\label{eq:2.18}
\end{equation}
where the dots denote terms which are of no interest to us here, and
\begin{equation}
C_F = \frac{N^2 -1}{2 N} \,.
\label{2.19}
\end{equation}

Applying the renormalization \eqn{eq:2.18} to the quark fields in the
operators $Q_i$ according to
\begin{equation}
\psi_{\rm B} = Z_\psi^{1/2} \psi \,,
\label{eq:2.20}
\end{equation}
one finds the matrices $\hZ_\psi^{(0)}$ and $\hZ_\psi^{(1)}$. They are
given by a single matrix $\hat{D}$
\begin{equation}
\hZ_\psi^{(0)} = -\hat{D},
\qquad
\hZ_\psi^{(1)} = \frac{3}{2}\, C_F \hat{D} \,.
\label{eq:2.21}
\end{equation}

The non--vanishing elements of $\hat{D}$ are
\begin{equation}
\begin{array}{ccccccccc}
\hat{D}(1,1) &=& \hat{D}(2,2) &=&  \frac{5}{9} \,, & & & & \\
\hat{D}(3,3)&=&\hat{D}(4,4)&=&\hat{D}(5,5) &=&\hat{D}(6,6)&=&\frac{1}{3} \,,\\
\hat{D}(3,9)&=&\hat{D}(4,10)&=&\hat{D}(5,7) &=&\hat{D}(6,8)&=&\frac{2}{9}\,, \\
\hat{D}(7,7)&=&\hat{D}(8,8)&=&\hat{D}(9,9) &=&\hat{D}(10,10)&=&\frac{4}{9}\,,\\
\hat{D}(7,5)&=&\hat{D}(8,6)&=&\hat{D}(9,3) &=&\hat{D}(10,4)&=&\frac{1}{9} \,.
\end{array}
\label{eq:2.22}
\end{equation}

\newsection{One-Loop Results}
We give here one--loop results in QED. The corresponding QCD expressions can
be found in section 3 of ref.~\cite{burasetal:92b}.

\subsection{Current-Current Contributions to $\gem$}
The non--vanishing contributions of diagrams $D_1$ -- $D_3$ of fig.~2 to the
matrix $\gem$ are given as follows
\begin{equation}
\begin{array}{ccccccccc}
\brackcc{\gem(1,1)}  &=& \brackcc{\gem(2,2)}   &=&
\brackcc{\gem(3,9)}  &=& \brackcc{\gem(4,10)}  &=& -\frac{8}{3}\,, \\
\svs
\brackcc{\gem(5,7)}  &=& \brackcc{\gem(6,8 )}  &=& \frac{8}{3}\,, & & & & \\
\svs
\brackcc{\gem(7,5)}  &=& \brackcc{\gem(7,7)}   &=&
\brackcc{\gem(8,6)}  &=& \brackcc{\gem(8,8)}   &=& \frac{4}{3}\,, \\
\svs
\brackcc{\gem(9,3)}  &=& \brackcc{\gem(9,9)}   &=&
\brackcc{\gem(10,4)} &=& \brackcc{\gem(10,10)} &=& -\frac{4}{3}\,.
\end{array}
\label{eq:3.1}
\end{equation}
These results already include the contributions from the wave function
renormalization.

\subsection{Penguin Contributions to $\gem$}
The contributions of diagrams $P_0^{(1)}$ and $P_0^{(2)}$ of fig.~3 to
the matrix $\gem$ have a very simple structure. The insertion of any
operator of \eqn{eq:2.1} into the diagrams of fig.~3 results always into
the sum of the operators $Q_7$ and $Q_9$ multiplied by an overall
factor. Denoting by
\begin{equation}
\bar{P} \; = \; \left(0,0,0,0,0,0,1,0,1,0\right)
\label{eq:3.2}
\end{equation}
the row vector in the space $(Q_1, \ldots, Q_{10})$, the
elements of $\gem$ coming from the diagrams of fig.~3 are as follows
\begin{equation}
\begin{array}{rclrcl}
\brackp{\gem(Q_1)} &=& \frac{16}{27} N \bar{P}\,, &
\brackp{\gem(Q_2)} &=& \frac{16}{27}   \bar{P}\,,  \\
\brackp{\gem(Q_3)} &=& \frac{16}{27} N \left(u-\frac{d}{2}-\frac{1}{N}
\right) \bar{P}\,, &
\brackp{\gem(Q_4)} &=& \frac{16}{27}   \left(u-\frac{d}{2}-N\right)\bar{P}\,,
\\
\brackp{\gem(Q_5)} &=& \frac{16}{27} N \left(u-\frac{d}{2}\right) \bar{P}\,, &
\brackp{\gem(Q_6)} &=& \frac{16}{27}   \left(u-\frac{d}{2}\right) \bar{P}\,,
\\
\brackp{\gem(Q_7)} &=& \frac{16}{27} N \left(u+\frac{d}{4}\right) \bar{P}\,, &
\brackp{\gem(Q_8)} &=& \frac{16}{27}   \left(u+\frac{d}{4}\right) \bar{P}\,,
\\
\brackp{\gem(Q_9)} &=& \frac{16}{27} N \left(u+\frac{d}{4}+\frac{1}{2N}
\right) \bar{P}\,, &
\brackp{\gem(Q_{10})}&=& \frac{16}{27} \left(u+\frac{d}{4}+\frac{N}{2}
\right) \bar{P}\,,
\end{array}
\label{eq:3.3}
\end{equation}
where $u$ and $d$ ($u+d=f$) denote the number of effective up-- and
down--quark flavours, respectively.

It should be stressed that these results do not depend on
renormalization scheme for $Q_{i}$ and are also valid for
$\widetilde{Q}_{i}$.

\subsection{General Comments on $\hat{r}$}

As already remarked in section 2.4, the finite terms $\rs$ and $\re$
depend on the renormalization scheme used. In the case of the NDR scheme
considered here, the penguin diagram contributions to these finite terms
may also depend on whether $Q_i$ or their Fierz conjugates
$\widetilde{Q}_i$ are inserted in the penguin diagrams. This observation
has already been made in ref.~\cite{burasetal:92b} and has been used to
avoid a direct evaluation of the two--loop diagrams involving
$Tr(\gf\gamma_\mu\gamma_\nu\gamma_\tau\gamma_\sigma)$, i.e.~the type 1
penguin insertions. Here we only state that whereas the results for
$(V-A)\otimes (V+A)$ operators $Q_5$--$Q_8$ do not depend on the form
used, this is no longer the case for the $(V-A)\otimes (V-A)$ operators
$Q_1$--$Q_4$, $Q_9$, and $Q_{10}$. We will return to this in section~5.1,
where a generalization of the method of ref.~\cite{burasetal:92b} to
the mixed QED--QCD case is presented.

\newsection{Current-Current Contributions to the $\ord(\aem\,\as)$
Anomalous Dimension Matrix $\gse$}

The calculation of current--current contributions to the $\ord(\aem\,\as)$
anomalous dimension matrix $\gse$ uses the singularities found in our
previous calculation of the $\ord(\as^2)$ contributions. Due to the
fact that certain diagrams present in the pure QCD case do not contribute
here and due to different colour and electric charge factors, the basic
structure of the $\ord(\aem\,\as)$ results is quite different from pure QCD.
In particular, the mixing between different operators can be divided
into three blocks $(Q_1,Q_2)$, $(Q_3,Q_4,Q_9,Q_{10})$ and
$(Q_5,Q_6,Q_7,Q_8)$ with no mixing between different blocks.

The results in this section include the contribution of the matrix
$\hZ_\psi^{(1)}$.

\subsection{$(V-A) \otimes (V-A)$ Operators}
The $2\times 2$ matrix describing  the mixing in the current--current
sector $(Q_1,Q_2)$ is given by
\begin{equation}
\left(
\begin{array}{cc}
8 N - \frac{22}{3}\frac{1}{N} & -\,\frac{2}{3} \svs \\
      \frac{25}{3}            & -N - \frac{22}{3}\frac{1}{N}
\end{array}
\right)\,.
\label{eq:4.3}
\end{equation}

The $4\times 4$ matrix describing  the mixing in the current--current
sector $(Q_3,Q_4,Q_9,Q_{10})$ is given by
\begin{equation}
\left(
\begin{array}{cccc}
\frac{6}{N} & -\,6 & 8 N - \frac{40}{3}\frac{1}{N} & \frac{16}{3} \svs \\
-\,3 & -\,3 N + \frac{6}{N} & \frac{34}{3} & 2 N - \frac{40}{3}\frac{1}{N}
\svs \\
 4 N - \frac{20}{3}\frac{1}{N} & \frac{8}{3} & 4 N - \frac{2}{3}\frac{1}{N} &
-\,\frac{10}{3} \\ \svs
\frac{17}{3} & N - \frac{20}{3}\frac{1}{N} & \frac{8}{3} &
-\,2 N - \frac{2}{3}\frac{1}{N} \svs
\end{array}
\right)\,.
\label{eq:4.4}
\end{equation}

\subsection{$(V-A) \otimes (V+A)$ Operators}

The $4\times 4$ matrix describing  the mixing in the current--current
sector $(Q_5,Q_6,Q_7,Q_8)$ is given by
\begin{equation}
\left(
\begin{array}{cccc}
-\,\frac{6}{N} & 6 & -\,8 N - \frac{16}{3}\frac{1}{N} & \frac{40}{3} \\
\svs
3 & 3 N - \frac{6}{N} & -\,2 & \frac{22}{3} N - \frac{16}{3}\frac{1}{N} \\
\svs
-\,4 N - \frac{8}{3}\frac{1}{N} & \frac{20}{3} & -\,4 N -
\frac{26}{3}\frac{1}{N} & \frac{38}{3} \\
\svs
-\,1 & \frac{11}{3} N - \frac{8}{3}\frac{1}{N} & 2 & \frac{20}{3} N -
\frac{26}{3}\frac{1}{N}
\end{array}
\right)\,.
\label{eq:4.5}
\end{equation}
We observe that the elements of both matrices grow at most as $N$ in the
Large--$N$ limit and consequently the anomalous dimension matrix
$\ord(\aem\,\as)$ resulting from current--current diagrams approaches a
constant $N$--independent matrix.

\newsection{Penguin Diagram Contributions to the $\ord(\aem\,\as)$
Anomalous Dimension Matrix $\gse$}

\subsection{General Structure}

The calculation of the penguin diagram contributions to $\gse$ can be
considerably simplified by first analyzing the general structure of
these contributions. The two--loop penguin diagrams are shown in fig.~5
where two types of insertions of a given operator, type 1 and type 2,
have to be considered. It is known that type 1 insertions are
problematic in the NDR scheme in which closed fermion loops involving $\gf$
can not be calculated unambiguously. Yet, as we have demonstrated in our
previous paper, the calculation of $\ord(\as^2)$ anomalous dimensions can be
reduced to the calculation of type 2 insertions only. Type 2 insertions
do not pose any problems. Indeed with the help of eq.~\eqn{eq:2.17}
with $a$ and $b$ standing this time for two different bases (see
eqs.~\eqn{eq:2.1} and \eqn{eq:2.2}) it is possible to relate the type
1 insertions to the type 2 insertions. This procedure can be
generalized to $\ord(\aem\,\as)$ so that also in this case explicit
calculations of closed fermion loops can be avoided. The procedure this
time is more complicated because the insertions of operators involving
$\bar u u$ have to be distinguished from the insertions of operators
involving $\bar d d$. We will now explain the procedure in some
detail.

In order to solve the problem of closed fermion loops involving $\gf$
in the NDR scheme, we have to consider four different bases of operators.
All four bases contain the penguin operators $Q_3$ -- $Q_{10}$ of
\eqn{eq:2.1} but differ in the current--current operators $Q_1$ and
$Q_2$. The latter are given as follows,

\leftline{\bf Basis A:}
\begin{equation}
Q_1^{(u)} \; = \; \left(\bar s_\alpha u_\beta  \right)_{V-A}
                  \left(\bar u_\beta  d_\alpha \right)_{V-A} \,,
\qquad
Q_2^{(u)} \; = \; \left(\bar s u \right)_{V-A}
                  \left(\bar u d \right)_{V-A} \,.
\label{eq:5.1}
\end{equation}

\leftline{\bf Basis B:}
\begin{equation}
Q_1^{(d)} \; = \; \left(\bar s_\alpha d_\beta  \right)_{V-A}
                  \left(\bar d_\beta  d_\alpha \right)_{V-A} \,,
\qquad
Q_2^{(d)} \; = \; \left(\bar s d \right)_{V-A}
                  \left(\bar d d \right)_{V-A} \,.
\label{eq:5.2}
\end{equation}

\leftline{\bf Basis C:}
\begin{equation}
\widetilde{Q}_1^{(u)} \; = \; \left(\bar s d \right)_{V-A}
                              \left(\bar u u \right)_{V-A} \,,
\qquad
\widetilde{Q}_2^{(u)} \; = \; \left(\bar s_\alpha d_\beta  \right)_{V-A}
                              \left(\bar u_\beta  u_\alpha \right)_{V-A} \,.
\label{eq:5.3}
\end{equation}

\leftline{\bf Basis D:}
\begin{equation}
\widetilde{Q}_1^{(d)} \; = \; \left(\bar s d \right)_{V-A}
                              \left(\bar d d \right)_{V-A} \,,
\qquad
\widetilde{Q}_2^{(d)} \; = \; \left(\bar s_\alpha d_\beta  \right)_{V-A}
                              \left(\bar d_\beta  d_\alpha \right)_{V-A} \,.
\label{eq:5.4}
\end{equation}

The basis A is the standard basis of \eqn{eq:2.1} and the basis B is
an auxiliary basis needed for the solution of the problem. The bases
C and D are simply Fierz conjugates of $Q_1$ and $Q_2$ in A and B,
respectively. Evidently,
\begin{equation}
Q_1^{(d)} \; = \; \widetilde{Q}_2^{(d)} \, ,
\qquad
Q_2^{(d)} \; = \; \widetilde{Q}_1^{(d)} \, .
\label{eq:5.5}
\end{equation}

Let us next denote by $[Q_i]_1$ and $[Q_i]_2$ the result of the type 1
and type 2 insertions of an operator $Q_i$, respectively. Then the
results for penguin contributions to the row entries of $\gse$ for
$Q_3$, $Q_4$, $Q_9$, and $Q_{10}$ can be written as follows
\begin{eqnarray}
{[Q_3]_{\rm p}} &=& u\,[\widetilde{Q}_1^{(u)}]_1 +
              d\,[\widetilde{Q}_1^{(d)}]_1 + 2\,[\widetilde{Q}_1^{(d)}]_2 \,,
              \label{eq:5.6} \\
\svs
{[Q_4]_{\rm p}} &=& u\,[\widetilde{Q}_2^{(u)}]_1 +
              d\,[\widetilde{Q}_2^{(d)}]_1 + 2\,[\widetilde{Q}_2^{(d)}]_2 \,,
              \label{eq:5.7} \\
\svs
{[Q_9]_{\rm p}} &=& u\,[\widetilde{Q}_1^{(u)}]_1 -
    \frac{d}{2}\,[\widetilde{Q}_1^{(d)}]_1 - [\widetilde{Q}_1^{(d)}]_2 \,,
              \label{eq:5.8} \\
\svs
{[Q_{10}]_{\rm p}} &=& u\,[\widetilde{Q}_2^{(u)}]_1 -
       \frac{d}{2}\,[\widetilde{Q}_2^{(d)}]_1 - [\widetilde{Q}_2^{(d)}]_2 \,.
                 \label{eq:5.9}
\end{eqnarray}

Evidently, in order to calculate these rows, the problem of closed
fermion loops has to be considered. $[Q_1]_{\rm p}$ and $[Q_2]_{\rm p}$
receive contributions only from type 2 insertions, i.e.~
\begin{equation}
[Q_1]_{\rm p} \; = \; [Q_1^{(u)}]_2 \,,
\qquad
[Q_2]_{\rm p} \; = \; [Q_2^{(u)}]_2 \,,
\label{eq:5.10}
\end{equation}
and do not pose any problems. The case of $(V-A)\otimes (V+A)$
operators will be discussed below.

There are six independent entries in eqs. \eqn{eq:5.6} -- \eqn{eq:5.9},
which have to be found in order to complete the calculation. We show
how this can be done in four steps:

\leftline{\bf Step 1:}
$[\widetilde{Q}_1^{(d)}]_2$ and $[\widetilde{Q}_2^{(d)}]_2$ can be calculated
without any problems. The results are given in section~5.2. Moreover
one finds the relation
\begin{equation}
[\widetilde{Q}_2^{(d)}]_2 \; = \; -\,\frac{1}{2}\,[Q_1^{(u)}]_2 \,.
\label{eq:5.11}
\end{equation}

\leftline{\bf Step 2:}
$\widetilde{Q}_1^{(u)}$ and $\widetilde{Q}_2^{(u)}$ receive contributions only
from type 1 insertions. This allows to find $[\widetilde{Q}_1^{(u)}]_1$ and
$[\widetilde{Q}_2^{(u)}]_1$ by comparing the two--loop anomalous dimension
matrices calculated in the bases C and A using the relation
\begin{equation}
\big(\gse\big)_p^{(C)} \; = \; \big(\gse\big)_p^{(A)} +
\left[ (\drs)_p, \gem \right] + \left[ (\dre)_p, \gs \right] \,,
\label{eq:5.12}
\end{equation}
where this time
\begin{equation}
\drs = \brackets{\rs}{C} - \brackets{\rs}{A} \,, \quad\quad
\dre = \brackets{\re}{C} - \brackets{\re}{A} \,.
\label{eq:5.12a}
\end{equation}

Because the insertions in the current--current diagrams are identical for
these two bases, only penguin diagram contributions to $\gse$ and
$\Delta\hat{r}_i$, $i=s,e$ enter this relation. On the other hand $\gem$
and $\gs$ are full one--loop matrices. A simple calculation of finite
terms in one--loop penguin diagrams gives
\begin{equation}
(\dre)_p \; = \; -\,\frac{8}{27} \left(
\begin{array}{c}
N \bar{P} \\ \bar{P} \\ 0 \\ \vdots \\ 0
\end{array}
\right) \,,
\qquad
(\drs)_p \; = \; -\,\frac{1}{3} \left(
\begin{array}{c}
 0 \\ P \\ 0 \\ \vdots \\ 0
\end{array}
\right) \,,
\label{eq:5.13}
\end{equation}
with $\bar{P}$ defined in \eqn{eq:3.2} and
\begin{equation}
P \; = \; \left( 0, \,0, \,-\,\frac{1}{N}, \,1, \,-\,\frac{1}{N}, \,1,
               \,0, \,0, \,0, \,0 \right)
\label{eq:5.13a}
\end{equation}
as in ref.~\cite{burasetal:92b}. The evaluation of \eqn{eq:5.13} does
not involve the dangerous traces with $\gf$.

\leftline{\bf Step 3:}
$[\widetilde{Q}_1^{(d)}]_1$ can be related to $[\widetilde{Q}_1^{(u)}]_1$ by
inspecting the diagrams of fig.~\ref{fig:5}. We find
\begin{equation}
[\widetilde{Q}_1^{(d)}]_1 \; = \; -\,\frac{1}{2}\,[\widetilde{Q}_1^{(u)}]_1\,,
\label{eq:5.14}
\end{equation}
and consequently $[\widetilde{Q}_1^{(d)}]_1$ can be found by using
\eqn{eq:5.14} and $[\widetilde{Q}_1^{(u)}]_1$ obtained in step 2.

\leftline{\bf Step 4:}
The calculation of $[\widetilde{Q}_2^{(d)}]_1$ is slightly more complicated
because the inspection of the diagrams of fig.~\ref{fig:5} does not
allow for a simple relation like \eqn{eq:5.14}. Since
$\widetilde{Q}_2^{(d)}$ receives contributions from both type 1 and 2
insertions we can write
\begin{equation}
[\widetilde{Q}_2^{(d)}]_1 \; = \; [\widetilde{Q}_2^{(d)}]_{\rm p} -
[\widetilde{Q}_2^{(d)}]_2 \,,
\label{eq:5.15}
\end{equation}
with the last entry calculated in step 1.

In order to find $[\widetilde{Q}_2^{(d)}]_{\rm p}$ we compare the two--loop
anomalous dimension matrices calculated in the bases D and B using
the relation
\begin{equation}
\big(\gse\big)_p^{(D)} \; = \; \big(\gse\big)_p^{(B)} +
\left[ (\drs)_{p}^{(d)}, (\gem)^{(d)} \right] +
\left[ (\dre)_{p}^{(d)}, (\gs)^{(d)} \right] \,,
\label{eq:5.16}
\end{equation}
where the index $d$ indicates that now the auxiliary bases D and B are
considered. We find first
\begin{equation}
(\dre)_{p}^{(d)} \; = \; \frac{4}{27}\,(N-1) \left(
\begin{array}{c}
 \bar{P} \\ -\bar{P} \\ 0 \\ \vdots \\ 0
\end{array}
\right) \,,
\qquad
(\drs)_{p}^{(d)} \; = \; \frac{1}{3} \left(
\begin{array}{c}
 P \\ -P \\ 0 \\ \vdots \\ 0
\end{array}
\right) \,.
\label{eq:5.17}
\end{equation}

The matrix $(\gs)^{(d)}$ differs from $\gs$ only in the first row which in
this case equals the second row. An explicit formula for $\gs$ can be
found in appendix A of ref.~\cite{burasetal:92b}. The matrix $(\gem)^{(d)}$
is given by eqs. \eqn{eq:3.1} -- \eqn{eq:3.3}, except for the following
changes,
\begin{equation}
\brackcc{\gem(1,1)}^{(d)} \; = \; \brackcc{\gem(2,2)}^{(d)} \; = \;
\frac{4}{3} \,,
\label{eq:5.18}
\end{equation}
\begin{equation}
\brackp{\gem(Q_1^{(d)})} \; = \; \brackp{\gem(Q_2^{(d)})} \; = \;
-\,\frac{8}{27}\,(N+1) \bar{P} \,.
\label{eq:5.19}
\end{equation}

In this way, the commutators in \eqn{eq:5.16} can be calculated. Since
moreover both types of insertions of $\widetilde{Q}_1^{(d)}$ have been
calculated in steps 1 and 3, the element $[\widetilde{Q}_2^{(d)}]_1$ can
finally be extracted from \eqn{eq:5.16} when in addition the relation
\eqn{eq:5.5} is used.

In summary, we have demonstrated that the penguin contributions to the
rows for $Q_3$, $Q_4$, $Q_9$, and $Q_{10}$ can be obtained by considering
only type 2 insertions.

The calculation of the rows for $Q_5$ -- $Q_8$ in the anomalous
dimension matrix is simplified by the fact that the Fierz symmetry is
preserved for these operators in the NDR scheme as already
discussed in section 3 and ref.~\cite{burasetal:92b}. Thus it is
sufficient to consider only the operators $\widetilde{Q}_5$ and
$\widetilde{Q}_6$ which receive only contributions from type 2 insertions.
This gives directly the $Q_5$ and $Q_6$ lines. In order to obtain the
$Q_7$ and $Q_8$ lines one has to replace $d$ by $-d/2$ in the $Q_5$ and
$Q_6$ lines, respectively.

\subsection{Results}
The singular terms in the diagrams of fig.~5 with type 2 insertions can
be found in tables 2--4 of ref.~\cite{burasetal:92b}. Using these
tables, including properly colour and electric charge factors, we can
find all the insertions necessary to calculate the full matrix
$(\gse)_p$ according to the procedure of the previous subsection.

We find
\begin{eqnarray}
\brackp{\gsendr(Q_1)} &=&
\frac{88}{243} \left[ N (Q_4 + Q_6) - (Q_3 + Q_5) \right] \nn \\
& &
+ \left[ \frac{8}{9} N^2 - \frac{64}{27} \right] Q_7
+ \left[ \frac{40}{27} N \right] Q_8  \nn \\
& &
+ \left[ \frac{8}{9} N^2 - \frac{80}{27} \right] Q_9
+ \left[ \frac{56}{27} N \right] Q_{10} \,,
\label{eq:5.22}
\end{eqnarray}
\begin{eqnarray}
\brackp{\gsendr(Q_2)} &=&
\frac{556}{243} \left[ (Q_4 + Q_6) - \frac{1}{N} (Q_3 + Q_5) \right] \nn \\
& &
+ \left[ -\frac{200}{243} N + \frac{1316}{243} \frac{1}{N} \right] Q_7
- \left[ \frac{124}{27} \right] Q_8  \nn \\
& &
+ \left[ -\frac{200}{243} N - \frac{1348}{243} \frac{1}{N} \right] Q_9
+ \left[ \frac{172}{27} \right] Q_{10} \,,
\label{eq:5.23}
\end{eqnarray}
\begin{eqnarray}
\brackp{\gsendr(\widetilde{Q}_5)} &=&
(u-d/2)\,\frac{136}{243} \left[ N (Q_4 + Q_6) - (Q_3 + Q_5) \right] \nn \\
& &
+ (u-d/2) \left[ \frac{8}{9} N^2 - \frac{112}{27} \right] Q_7
+ \left[ (u-d/2)\,\frac{88}{27} N \right] Q_8  \nn \\
& &
+ (u-d/2) \left[ \frac{8}{9} N^2 - \frac{32}{27} \right] Q_9
+ \left[ (u-d/2)\,\frac{8}{27} N \right] Q_{10} \,,
\label{eq:5.24}
\end{eqnarray}
\begin{eqnarray}
\brackp{\gsendr(\widetilde{Q}_6)} &=&
\left[ \frac{8}{9} (u+d/4) + \frac{532}{243} (u-d/2) \right]
\left[ (Q_4 + Q_6) - \frac{1}{N} (Q_3 + Q_5) \right] \nn \\
& &
+ \left[ (u-d/2) \left(-\frac{64}{27} N - \frac{56}{9} \frac{1}{N}\right) +
f \left( \frac{136}{243} N + \frac{260}{243} \frac{1}{N}\right) \right] Q_7 \nn
\\
& &
+ \left[ \frac{232}{27} (u-d/2) - \frac{44}{27} f \right] Q_8  \nn \\
& &
+ \left[ (u-d/2) \left(-\frac{64}{27} N + \frac{200}{27} \frac{1}{N}\right) +
f \left( \frac{136}{243} N - \frac{100}{243} \frac{1}{N} \right)\right] Q_9 \nn
\\
& &
+ \left[ -\frac{136}{27} (u-d/2) - \frac{4}{27} f \right] Q_{10} \,,
\label{eq:5.25}
\end{eqnarray}
and
\begin{eqnarray}
\left[ \gsendr(\widetilde{Q}_1^{(d)}) \right]_2 &=&
-\,\frac{116}{243} \left[ (Q_4 + Q_6) - \frac{1}{N} (Q_3 + Q_5) \right] \nn \\
& &
+ \left[ \frac{232}{243} N - \frac{520}{243} \frac{1}{N} \right] Q_7
+ \left[ \frac{32}{27} \right] Q_8  \nn \\
& &
+ \left[ \frac{232}{243} N + \frac{920}{243} \frac{1}{N} \right] Q_9
- \left[ \frac{128}{27} \right] Q_{10} \,,
\label{eq:5.26}
\end{eqnarray}
\begin{equation}
\left[ \gsendr(\widetilde{Q}_2^{(d)}) \right]_2 \; = \; -\,\frac{1}{2}\,
\brackp{\gsendr(Q_1)} \,,
\label{eq:5.27}
\end{equation}
where we have indicated that these results have been obtained in the NDR
scheme.
We observe that $\brackp{\gem}$ (see eq.~\eqn{eq:3.3}) and $\gse$ contain
elements growing like $N$  and $N^2$, respectively. Consequently, the
corresponding elements in the one--loop $\ord(\aem)$ and two--loop
$\ord(\aem\,\as)$ matrices grow like $N$ in the Large--$N$ limit.

\newsection{Full two--loop Anomalous Dimension Matrix $\gss$}

\subsection{Basic Result of this Paper}

Adding the current--current and penguin contributions to $\gse$ found in
sections~4 and 5, respectively, we obtain the complete $\ord(\aem\,\as)$
anomalous dimension matrix $\gsendr$.

We first note that 16 elements of this matrix vanish. These are the
entries $[\gsendr]_{ij}$ with $i=3,\ldots,10;\/ j=1,2$.

The remaining 84 elements are non--vanishing. These entries of
$\gsendr$ are given in table~\ref{tab:1}. For phenomenological
applications, we need only the results with $N=3$.  We give them in
appendix~C for an arbitrary number of flavours.

Let us just comment on certain features of the matrix $\gsendr$.
\begin{itemize}
\item A comparison of $\gem$ and $\gsendr$ shows that the impact of
QCD is to fill out most of the zero entries present in $\gem$.
\item
The corrections to non--vanishing elements in $\gem$ are moderate. For
$N=3$ and $u=d=2$ the largest corrections are found in elemets $(7,5)$,
$(8,6)$ and $(4,7)$. For $\as/4\pi \approx 0.01$ they amount to
$10-15\;\%$ corrections. Thus the corrections are certainly smaller than
in the case of $\gssndr$ evaluated in ref.~\cite{burasetal:92b}.
Similar comments apply to elements which become non--zero at
$\ord(\aem\,\as)$. They are smaller than the corresponding terms in the
matrix $\gss$.
\item
It is needless to say that all these comments are specific to the NDR scheme
and the true size of the next--to--leading order corrections can only be
assessed after a full renormalization group analysis has been performed.
We will return to this question in ref.~\cite{burasetal:92d}.
\end{itemize}

As in the case of the ${\cal O}(\alpha_{s}^{2})$ corrections, on general
grounds the coefficient of the $1/\varepsilon^{2}$--divergences at
${\cal O}(\alpha\alpha_{s})$ in the unrenormalized Green function,
$\hat\Gamma^{\,e}_{22}$, is entirely given in terms of quantities
calculable at one--loop,
\begin{equation}
\hat\Gamma^{\,e}_{22} \; = \; -\,\frac{1}{8}\,\left[\,\gem\gs+\gs\gem\,\right]
-\frac{1}{2}\,\gs\hat Z^{(0)}_{\psi}-a_{1}\gem-2\,a_{1}\hat Z^{(0)}_{\psi}
\label{eq:6.a}
\end{equation}
with $a_1$ and $\gs$ given in ref.~\cite{burasetal:92b}.
The explicit calculation shows that this relation is indeed satisfied,
which constitutes a check of our calculation. For details, the reader
is referred to sect.~6.3 of ref.~\cite{burasetal:92b}.

\subsection{Comments on the HV Scheme}

Having the two--loop anomalous dimension matrix $\gsendr$ at hand, it
is a simple matter to obtain this matrix in any other scheme by using
the relation \eqn{eq:2.17}. As an example, we consider here the HV
scheme for which $\gsshv$ has been explicitly given in
ref.~\cite{burasetal:92b}.

In order to find $\gsehv$, the simplest method is to use
\begin{equation}
\gsehv \; = \; \gsendr + \left[ \drs, \gem \right] +
                         \left[ \dre, \gs  \right] \,,
\label{eq:6.1}
\end{equation}
where this time
\begin{equation}
\drs \equiv (\rs)_{\rm HV} - (\rs)_{\rm NDR} \, , \qquad
\dre \equiv (\re)_{\rm HV} - (\re)_{\rm NDR} \, .
\label{eq:6.2}
\end{equation}

The result for $\drs$ can be found in sections~3.3 and 3.4 of
ref.~\cite{burasetal:92b}. Here we give in addition the result for
$\dre$. As in the case of $\drs$ it is convenient to separate the
contributions to $\dre$ into current--current and penguin contributions
obtained from finite terms in figs.~\ref{fig:2} and \ref{fig:3},
respectively.

The non--vanishing elements in $\brackcc{\dre}$ are as follows:
\begin{equation}
\begin{array}{lclclclcr}
 \brackcc{ \dre(1,1) } &=&
 \brackcc{ \dre(2,2) } &=&
-\brackcc{ \dre(9,3) } &=&
-\brackcc{ \dre(10,4) } &=& \frac{2}{9} \, , \\
\svs
\brackcc{ \dre(3,3) } &=&
\brackcc{ \dre(4,4) } &=&
\brackcc{ \dre(5,5) } &=&
\brackcc{ \dre(6,6) } &=& \frac{2}{3} \, , \\
\svs
\brackcc{ \dre(3,9) } &=&
\brackcc{ \dre(4,10) } &=&
\brackcc{ \dre(7,7) } &=&
\brackcc{ \dre(8,8) } &=& -\frac{4}{9} \, , \\
\svs
\brackcc{ \dre(9,9) } &=&
\brackcc{ \dre(10,10) } &=& \frac{4}{9} \, ,
& & & & \\
  \brackcc{ \dre(5,7) } &=&
  \brackcc{ \dre(6,8) } &=&
2 \brackcc{ \dre(7,5) } &=&
2 \brackcc{ \dre(8,6) } &=& -\frac{20}{9} \, . \\
\label{eq:6.3}
\end{array}
\end{equation}

For the penguin contributions one gets:
\begin{equation}
\begin{array}{lclclclcl}
  \brackp{ \dre(Q_1)    } &=&
 -\brackp{ \dre(Q_4)    } &=&
2 \brackp{ \dre(Q_{10}) } &=& -\frac{8}{27} N \bar{P} \, , & & \\
\svs
  \brackp{ \dre(Q_2) } &=&
 -\brackp{ \dre(Q_3) } &=&
2 \brackp{ \dre(Q_9) } &=& -\frac{8}{27} \bar{P} \, , & & \\
\svs
\brackp{ \dre(Q_5) } &=&
\brackp{ \dre(Q_6) } &=&
\brackp{ \dre(Q_7) } &=&
\brackp{ \dre(Q_8) } &=& 0 \, . \\
\label{eq:6.4}
\end{array}
\end{equation}

With the help of eqs.~\eqn{eq:6.1}--\eqn{eq:6.4} and $\drs$ of
ref.~\cite{burasetal:92b}, one can find an explicit expression for
$\gsehv$. We do not give it here since due to the complexity of this
matrix it is anyway better to make the transformation \eqn{eq:6.1} by
computer.

\newsection{Summary}
We have presented the details and the explicit
results of the calculation of the $10 \times 10$ two--loop anomalous
dimension matrix $\ord(\aem\,\as)$ involving current--current,
QCD--penguin and electroweak penguin operators.

Performing the calculation in the simplest renormalization scheme
with anticommuting $\gf$ (NDR) we have demonstrated that a direct
evaluation of penguin diagrams with $\gf$ appearing in closed fermion
loops can be avoided by studying simultaneously four different bases of
operators. In this way an unambiguous result for $\gse$ in the NDR
scheme could be obtained.

This analysis completes our extensive calculations of the anomalous
dimension matrix $\hg$ in eq.~\eqn{eq:1.1} which we have presented in
\cite{burasweisz:90,burasetal:92a,burasetal:92b} and in the present
paper.
This matrix constitutes an important ingredient of any analysis of
non--leptonic weak decays which goes beyond the leading logarithmic
approximation.
The full next--to--leading renormalization group analysis of the
Wilson coefficient functions of operators $Q_i$ has been just completed
and  the details of this work together with phenomenological
implications can be found in ref.~\cite{burasetal:92d}.


\vskip 1cm
\begin{center}
{\Large\bf Acknowledgement}
\end{center}
\noindent
We would like to thank many colleagues for a continuous encouragement
during this calculations.  We are especially grateful to Peter Weisz
for the most pleasent collaboration in the QCD part of this project and
for his continued interest and many fruitful discussions on the present
paper. One of us (A.J.B.) would like to thank Bill Bardeen, David
Broadhurst and Dirk Kreimer for very interesting discussions.
M.E.L.~is grateful to Stefan Herrlich for a copy of his program {\tt
feynd} for drawing the Feynman diagrams in the figures and to Gerhard
Buchalla for stimulating discussions.


\newpage
\appendix{\LARGE\bf\noindent Appendices}

\newsection{One--Loop Anomalous Dimension Matrix $\gem$}

\bigskip

$ \gem =
\left(
\begin{array}{cccccccccc}
-{8\over 3} & 0 & 0 & 0 & 0 & 0 & {{16\,N}\over {27}} & 0 & {{16\,N}\over\
  {27}} & 0 \\ \svs
0 & -{8\over 3} & 0 & 0 & 0 & 0 & {{16}\over {27}} & 0 & {{16}\over {27}} & 0\
  \\ \svs
0 & 0 & 0 & 0 & 0 & 0 & -{{16}\over {27}} + {{16\,N\,\left( u-d/2\
  \right) }\over {27}} & 0 & -{{88}\over {27}} + {{16\,N\,\left(
  u-d/2 \right) }\over {27}} & 0 \\ \svs
0 & 0 & 0 & 0 & 0 & 0 & {{-16\,N}\over {27}} + {{16\,\left( u-d/2\
  \right) }\over {27}} & 0 & {{-16\,N}\over {27}} + {{16\,\left(
  u-d/2 \right) }\over {27}} & -{8\over 3} \\ \svs
0 & 0 & 0 & 0 & 0 & 0 & {8\over 3} + {{16\,N\,\left( u-d/2\
  \right) }\over {27}} & 0 & {{16\,N\,\left( u-d/2 \right) }\over\
  {27}} & 0 \\ \svs
0 & 0 & 0 & 0 & 0 & 0 & {{16\,\left( u-d/2 \right) }\over {27}} &\
  {8\over 3} & {{16\,\left( u-d/2 \right) }\over {27}} & 0 \\ \svs
0 & 0 & 0 & 0 & {4\over 3} & 0 & {4\over 3} + {{16\,N\,\left( u+d/4\
  \right) }\over {27}} & 0 & {{16\,N\,\left( u+d/4 \right) }\over\
  {27}} & 0 \\ \svs
0 & 0 & 0 & 0 & 0 & {4\over 3} & {{16\,\left( u+d/4 \right) }\over\
  {27}} & {4\over 3} & {{16\,\left( u+d/4 \right) }\over {27}} & 0\
  \\ \svs
0 & 0 & -{4\over 3} & 0 & 0 & 0 & {8\over {27}} + {{16\,N\,\left(
  u+d/4 \right) }\over {27}} & 0 & -{{28}\over {27}} + {{16\,N\,\left(
  u+d/4 \right) }\over {27}} & 0 \\ \svs
0 & 0 & 0 & -{4\over 3} & 0 & 0 & {{8\,N}\over {27}} + {{16\,\left(
  u+d/4 \right) }\over {27}} & 0 & {{8\,N}\over {27}} + {{16\,\left(
  u+d/4 \right) }\over {27}} & -{4\over 3}
\end{array}
\right) $


\bigskip


\clearpage
\newsection{Table for Two--Loop Anomalous Dimension Matrix $\gse$
in the NDR Scheme}
\begin{table}[ht]
\caption[]{Full QCD-QED Anomalous Dimension Matrix $(\gse)_{ij}$
for the NDR scheme (with vanishing entries omitted tacitly).
\label{tab:1}
}
\begin{center}
\begin{tabular}{|c|c||c|c|}
\hline
$(i,j)$ & & $(i,j)$ &  \svs \\
\hline\hline
$(1,1)$ & ${{-22}\over {3\,N}} + 8\,N$ &
$(1,2)$ & $-{2\over 3}$ \svs \\
$(1,3)$ & $-{{88}\over {243}}$ &
$(1,4)$ & ${{88\,N}\over {243}}$ \svs \\
$(1,5)$ & $-{{88}\over {243}}$ &
$(1,6)$ & ${{88\,N}\over {243}}$ \svs \\
$(1,7)$ & $-{{64}\over {27}} + {{8\,{N^2}}\over 9}$ &
$(1,8)$ & ${{40\,N}\over {27}}$ \svs \\
$(1,9)$ & $-{{80}\over {27}} + {{8\,{N^2}}\over 9}$ &
$(1,10)$ & ${{56\,N}\over {27}}$ \svs \\
\hline
$(2,1)$ & ${{25}\over 3}$ &
$(2,2)$ & ${{-22}\over {3\,N}} - N$ \svs \\
$(2,3)$ & ${{-556}\over {243\,N}}$ &
$(2,4)$ & ${{556}\over {243}}$ \svs \\
$(2,5)$ & ${{-556}\over {243\,N}}$ &
$(2,6)$ & ${{556}\over {243}}$ \svs \\
$(2,7)$ & ${{1316}\over {243\,N}} - {{200\,N}\over {243}}$ &
$(2,8)$ & $-{{124}\over {27}}$ \svs \\
$(2,9)$ & ${{-1348}\over {243\,N}} - {{200\,N}\over {243}}$ &
$(2,10)$ & ${{172}\over {27}}$ \svs \\
\hline
$(3,3)$ & ${{1690}\over {243\,N}} - {{136\,\left( u - d/2 \right) }\over\
  {243}}$ &
$(3,4)$ & $-{{1690}\over {243}} + {{136\,N\,\left( u - d/2 \right) }\over\
  {243}}$ \svs \\
$(3,5)$ & ${{232}\over {243\,N}} - {{136\,\left( u - d/2 \right) }\over\
  {243}}$ &
$(3,6)$ & $-{{232}\over {243}} + {{136\,N\,\left( u - d/2 \right) }\over\
  {243}}$ \svs \\
$(3,7)$ & ${{-1040}\over {243\,N}} + {{464\,N}\over {243}} - {{112\,\left( u\
  - d/2 \right) }\over {27}} + {{8\,{N^2}\,\left( u - d/2 \right) }\over 9}$\
  &
$(3,8)$ & ${{64}\over {27}} + {{88\,N\,\left( u - d/2 \right) }\over {27}}$
  \svs \\
$(3,9)$ & ${{-1400}\over {243\,N}} + {{2408\,N}\over {243}} - {{32\,\left( u\
  - d/2 \right) }\over {27}} + {{8\,{N^2}\,\left( u - d/2 \right) }\over 9}$\
  &
$(3,10)$ & $-{{112}\over {27}} + {{8\,N\,\left( u - d/2 \right) }\over {27}}$\
  \svs \\
\hline
$(4,3)$ & $-{{641}\over {243}} + {{{{-388\,u}\over {243}} + {{32\,d}\over\
  {243}}}\over N}$ &
$(4,4)$ & ${6\over N} - {{817\,N}\over {243}} + {{388\,u}\over {243}} -\
  {{32\,d}\over {243}}$ \svs \\
$(4,5)$ & ${{88}\over {243}} + {{{{-388\,u}\over {243}} + {{32\,d}\over\
  {243}}}\over N}$ &
$(4,6)$ & ${{-88\,N}\over {243}} + {{388\,u}\over {243}} - {{32\,d}\over\
  {243}}$ \svs \\
$(4,7)$ & ${{64}\over {27}} - {{8\,{N^2}}\over 9} + N\,\left( {{280\,u}\over\
  {243}} + {{64\,d}\over {243}} \right)  + {{{{620\,u}\over {243}} +\
  {{80\,d}\over {243}}}\over N}$ &
$(4,8)$ & ${{-40\,N}\over {27}} - {{100\,u}\over {27}} - {{16\,d}\over {27}}$\
  \svs \\
$(4,9)$ & ${{386}\over {27}} - {{8\,{N^2}}\over 9} + N\,\left( {{280\,u}\over\
  {243}} + {{64\,d}\over {243}} \right)  + {{{{-1612\,u}\over {243}} +\
  {{656\,d}\over {243}}}\over N}$ &
$(4,10)$ & ${{-40}\over {3\,N}} - {{2\,N}\over {27}} + {{148\,u}\over {27}} -\
  {{80\,d}\over {27}}$ \svs \\
\hline
$(5,3)$ & ${{-136\,\left( u - d/2 \right) }\over {243}}$ &
$(5,4)$ & ${{136\,N\,\left( u - d/2 \right) }\over {243}}$ \svs \\
$(5,5)$ & ${{-6}\over N} - {{136\,\left( u - d/2 \right) }\over {243}}$ &
$(5,6)$ & $6 + {{136\,N\,\left( u - d/2 \right) }\over {243}}$ \svs \\
$(5,7)$ & ${{-16}\over {3\,N}} - 8\,N - {{112\,\left( u - d/2 \right) }\over\
  {27}} + {{8\,{N^2}\,\left( u - d/2 \right) }\over 9}$ &
$(5,8)$ & ${{40}\over 3} + {{88\,N\,\left( u - d/2 \right) }\over {27}}$
  \svs \\
$(5,9)$ & ${{-32\,\left( u - d/2 \right) }\over {27}} + {{8\,{N^2}\,\left( u\
  - d/2 \right) }\over 9}$ &
$(5,10)$ & ${{8\,N\,\left( u - d/2 \right) }\over {27}}$ \svs \\
\hline
\end{tabular}
\end{center}
\end{table}

\addtocounter{table}{-1}
\begin{table}[ht]
\caption[]{Full QCD-QED Anomalous Dimension Matrix $(\gse)_{ij}$
for the NDR scheme (continued; with vanishing entries omitted tacitly).
}
\begin{center}
\begin{tabular}{|c|c||c|c|}
\hline
$(i,j)$ & & $(i,j)$ & \svs \\
\hline\hline
$(6,3)$ & ${{{{-748\,u}\over {243}} + {{212\,d}\over {243}}}\over N}$ &
$(6,4)$ & ${{748\,u}\over {243}} - {{212\,d}\over {243}}$ \svs \\
$(6,5)$ & $3 + {{{{-748\,u}\over {243}} + {{212\,d}\over {243}}}\over N}$ &
$(6,6)$ & ${{-6}\over N} + 3\,N + {{748\,u}\over {243}} - {{212\,d}\over\
  {243}}$ \svs \\
$(6,7)$ & $-2 + N\,\left( {{-440\,u}\over {243}} + {{424\,d}\over {243}}\
  \right)  + {{{{-1252\,u}\over {243}} + {{1016\,d}\over {243}}}\over N}$ &
$(6,8)$ & ${{-16}\over {3\,N}} + {{22\,N}\over 3} + {{188\,u}\over {27}} -\
  {{160\,d}\over {27}}$ \svs \\
$(6,9)$ & ${{{{1700\,u}\over {243}} - {{1000\,d}\over {243}}}\over N} +\
  N\,\left( {{-440\,u}\over {243}} + {{424\,d}\over {243}} \right) $ &
$(6,10)$ & ${{-140\,u}\over {27}} + {{64\,d}\over {27}}$ \svs \\
\hline
$(7,3)$ & ${{-136\,\left( u + d/4 \right) }\over {243}}$ &
$(7,4)$ & ${{136\,N\,\left( u + d/4 \right) }\over {243}}$ \svs \\
$(7,5)$ & ${{-8}\over {3\,N}} - 4\,N - {{136\,\left( u + d/4 \right) }\over\
  {243}}$ &
$(7,6)$ & ${{20}\over 3} + {{136\,N\,\left( u + d/4 \right) }\over {243}}$
  \svs \\
$(7,7)$ & ${{-26}\over {3\,N}} - 4\,N - {{112\,\left( u + d/4 \right) }\over\
  {27}} + {{8\,{N^2}\,\left( u + d/4 \right) }\over 9}$ &
$(7,8)$ & ${{38}\over 3} + {{88\,N\,\left( u + d/4 \right) }\over {27}}$
  \svs \\
$(7,9)$ & ${{-32\,\left( u + d/4 \right) }\over {27}} + {{8\,{N^2}\,\left( u\
  + d/4 \right) }\over 9}$ &
$(7,10)$ & ${{8\,N\,\left( u + d/4 \right) }\over {27}}$ \svs \\
\hline
$(8,3)$ & ${{{{-748\,u}\over {243}} - {{106\,d}\over {243}}}\over N}$ &
$(8,4)$ & ${{748\,u}\over {243}} + {{106\,d}\over {243}}$ \svs \\
$(8,5)$ & $-1 + {{{{-748\,u}\over {243}} - {{106\,d}\over {243}}}\over N}$ &
$(8,6)$ & ${{-8}\over {3\,N}} + {{11\,N}\over 3} + {{748\,u}\over {243}} +\
  {{106\,d}\over {243}}$ \svs \\
$(8,7)$ & $2 + {{{{-1252\,u}\over {243}} - {{508\,d}\over {243}}}\over N} +\
  N\,\left( {{-440\,u}\over {243}} - {{212\,d}\over {243}} \right) $ &
$(8,8)$ & ${{-26}\over {3\,N}} + {{20\,N}\over 3} + {{188\,u}\over {27}} +\
  {{80\,d}\over {27}}$ \svs \\
$(8,9)$ & $N\,\left( {{-440\,u}\over {243}} - {{212\,d}\over {243}} \right) \
  + {{{{1700\,u}\over {243}} + {{500\,d}\over {243}}}\over N}$ &
$(8,10)$ & ${{-140\,u}\over {27}} - {{32\,d}\over {27}}$ \svs \\
\hline
$(9,3)$ & ${{-1736}\over {243\,N}} + 4\,N - {{136\,\left( u + d/4 \right)\
  }\over {243}}$ &
$(9,4)$ & ${{764}\over {243}} + {{136\,N\,\left( u + d/4 \right) }\over\
  {243}}$ \svs \\
$(9,5)$ & ${{-116}\over {243\,N}} - {{136\,\left( u + d/4 \right) }\over\
  {243}}$ &
$(9,6)$ & ${{116}\over {243}} + {{136\,N\,\left( u + d/4 \right) }\over\
  {243}}$ \svs \\
$(9,7)$ & ${{520}\over {243\,N}} - {{232\,N}\over {243}} - {{112\,\left( u +\
  d/4 \right) }\over {27}} + {{8\,{N^2}\,\left( u + d/4 \right) }\over 9}$ &
$(9,8)$ & $-{{32}\over {27}} + {{88\,N\,\left( u + d/4 \right) }\over {27}}$\
  \svs \\
$(9,9)$ & ${{-1082}\over {243\,N}} + {{740\,N}\over {243}} - {{32\,\left( u +\
  d/4 \right) }\over {27}} + {{8\,{N^2}\,\left( u + d/4 \right) }\over 9}$ &
$(9,10)$ & ${{38}\over {27}} + {{8\,N\,\left( u + d/4 \right) }\over {27}}$
  \svs \\
\hline
$(10,3)$ & ${{1333}\over {243}} + {{{{-388\,u}\over {243}} - {{16\,d}\over\
  {243}}}\over N}$ &
$(10,4)$ & ${{-20}\over {3\,N}} + {{287\,N}\over {243}} + {{388\,u}\over\
  {243}} + {{16\,d}\over {243}}$ \svs \\
$(10,5)$ & $-{{44}\over {243}} + {{{{-388\,u}\over {243}} - {{16\,d}\over\
  {243}}}\over N}$ &
$(10,6)$ & ${{44\,N}\over {243}} + {{388\,u}\over {243}} + {{16\,d}\over\
  {243}}$ \svs \\
$(10,7)$ & $-{{32}\over {27}} + {{4\,{N^2}}\over 9} + {{{{620\,u}\over {243}}\
  - {{40\,d}\over {243}}}\over N} + N\,\left( {{280\,u}\over {243}} -\
  {{32\,d}\over {243}} \right) $ &
$(10,8)$ & ${{20\,N}\over {27}} - {{100\,u}\over {27}} + {{8\,d}\over {27}}$\
  \svs \\
$(10,9)$ & ${{32}\over {27}} + {{4\,{N^2}}\over 9} + {{{{-1612\,u}\over\
  {243}} - {{328\,d}\over {243}}}\over N} + N\,\left( {{280\,u}\over {243}} -\
  {{32\,d}\over {243}} \right) $ &
$(10,10)$ & ${{-2}\over {3\,N}} - {{26\,N}\over {27}} + {{148\,u}\over {27}}\
  + {{40\,d}\over {27}}$ \svs \\
\hline
\end{tabular}
\end{center}
\end{table}

\clearpage
\newsection{Two--Loop QCD--QED Anomalous Dimension Matrix $\gse$ for
$N=3$ in the NDR Scheme}
\begin{displaymath}
\hskip -2cm
\gsendr\bigl|_{N=3} =
\left(
\begin{array}{ccccc}
{{194}\over 9} & -{2\over 3} & -{{88}\over {243}} & {{88}\over {81}} &\
  -{{88}\over {243}} \msvs \\
{{25}\over 3} & -{{49}\over 9} & -{{556}\over {729}} & {{556}\over {243}} &\
  -{{556}\over {729}} \msvs \\
0 & 0 & {{1690}\over {729}} - {{136\,\left( u - d/2 \right) }\over {243}} &\
  -{{1690}\over {243}} + {{136\,\left( u - d/2 \right) }\over {81}} &\
  {{232}\over {729}} - {{136\,\left( u - d/2 \right) }\over {243}} \msvs \\
0 & 0 & -{{641}\over {243}} - {{388\,u}\over {729}} + {{32\,d}\over {729}} &\
  -{{655}\over {81}} + {{388\,u}\over {243}} - {{32\,d}\over {243}} &\
  {{88}\over {243}} - {{388\,u}\over {729}} + {{32\,d}\over {729}} \msvs \\
0 & 0 & {{-136\,\left( u - d/2 \right) }\over {243}} & {{136\,\left( u - d/2\
  \right) }\over {81}} & -2 - {{136\,\left( u - d/2 \right) }\over {243}} \msvs
\\
0 & 0 & {{-748\,u}\over {729}} + {{212\,d}\over {729}} & {{748\,u}\over\
  {243}} - {{212\,d}\over {243}} & 3 - {{748\,u}\over {729}} + {{212\,d}\over\
  {729}} \msvs \\
0 & 0 & {{-136\,\left( u + d/4 \right) }\over {243}} & {{136\,\left( u + d/4\
  \right) }\over {81}} & -{{116}\over 9} - {{136\,\left( u + d/4 \right)\
  }\over {243}} \msvs \\
0 & 0 & {{-748\,u}\over {729}} - {{106\,d}\over {729}} & {{748\,u}\over\
  {243}} + {{106\,d}\over {243}} & -1 - {{748\,u}\over {729}} -\
  {{106\,d}\over {729}} \msvs \\
0 & 0 & {{7012}\over {729}} - {{136\,\left( u + d/4 \right) }\over {243}} &\
  {{764}\over {243}} + {{136\,\left( u + d/4 \right) }\over {81}} &\
  -{{116}\over {729}} - {{136\,\left( u + d/4 \right) }\over {243}} \msvs \\
0 & 0 & {{1333}\over {243}} - {{388\,u}\over {729}} - {{16\,d}\over {729}} &\
  {{107}\over {81}} + {{388\,u}\over {243}} + {{16\,d}\over {243}} &\
  -{{44}\over {243}} - {{388\,u}\over {729}} - {{16\,d}\over {729}} \msvs
\end{array}
\right.
\end{displaymath}

\begin{displaymath}
\hskip -1cm
\left.
\begin{array}{ccccc}
{{88}\over {81}} & {{152}\over {27}} & {{40}\over 9} & {{136}\over {27}} &\
  {{56}\over 9} \msvs \\
{{556}\over {243}} & -{{484}\over {729}} & -{{124}\over {27}} & -{{3148}\over\
  {729}} & {{172}\over {27}} \msvs \\
-{{232}\over {243}} + {{136\,\left( u - d/2 \right) }\over {81}} &\
  {{3136}\over {729}} + {{104\,\left( u - d/2 \right) }\over {27}} &\
  {{64}\over {27}} + {{88\,\left( u - d/2 \right) }\over 9} & {{20272}\over\
  {729}} + {{184\,\left( u - d/2 \right) }\over {27}} & -{{112}\over {27}} +\
  {{8\,\left( u - d/2 \right) }\over 9} \msvs \\
-{{88}\over {81}} + {{388\,u}\over {243}} - {{32\,d}\over {243}} &\
  -{{152}\over {27}} + {{3140\,u}\over {729}} + {{656\,d}\over {729}} &\
  -{{40}\over 9} - {{100\,u}\over {27}} - {{16\,d}\over {27}} & {{170}\over\
  {27}} + {{908\,u}\over {729}} + {{1232\,d}\over {729}} & -{{14}\over 3} +\
  {{148\,u}\over {27}} - {{80\,d}\over {27}} \msvs \\
6 + {{136\,\left( u - d/2 \right) }\over {81}} & -{{232}\over 9} +\
  {{104\,\left( u - d/2 \right) }\over {27}} & {{40}\over 3} + {{88\,\left( u\
  - d/2 \right) }\over 9} & {{184\,\left( u - d/2 \right) }\over {27}} &\
  {{8\,\left( u - d/2 \right) }\over 9} \msvs \\
7 + {{748\,u}\over {243}} - {{212\,d}\over {243}} & -2 - {{5212\,u}\over\
  {729}} + {{4832\,d}\over {729}} & {{182}\over 9} + {{188\,u}\over {27}} -\
  {{160\,d}\over {27}} & {{-2260\,u}\over {729}} + {{2816\,d}\over {729}} &\
  {{-140\,u}\over {27}} + {{64\,d}\over {27}} \msvs \\
{{20}\over 3} + {{136\,\left( u + d/4 \right) }\over {81}} & -{{134}\over 9}\
  + {{104\,\left( u + d/4 \right) }\over {27}} & {{38}\over 3} + {{88\,\left(\
  u + d/4 \right) }\over 9} & {{184\,\left( u + d/4 \right) }\over {27}} &\
  {{8\,\left( u + d/4 \right) }\over 9} \msvs \\
{{91}\over 9} + {{748\,u}\over {243}} + {{106\,d}\over {243}} & 2 -\
  {{5212\,u}\over {729}} - {{2416\,d}\over {729}} & {{154}\over 9} +\
  {{188\,u}\over {27}} + {{80\,d}\over {27}} & {{-2260\,u}\over {729}} -\
  {{1408\,d}\over {729}} & {{-140\,u}\over {27}} - {{32\,d}\over {27}} \msvs \\
{{116}\over {243}} + {{136\,\left( u + d/4 \right) }\over {81}} &\
  -{{1568}\over {729}} + {{104\,\left( u + d/4 \right) }\over {27}} &\
  -{{32}\over {27}} + {{88\,\left( u + d/4 \right) }\over 9} & {{5578}\over\
  {729}} + {{184\,\left( u + d/4 \right) }\over {27}} & {{38}\over {27}} +\
  {{8\,\left( u + d/4 \right) }\over 9} \msvs \\
{{44}\over {81}} + {{388\,u}\over {243}} + {{16\,d}\over {243}} & {{76}\over\
  {27}} + {{3140\,u}\over {729}} - {{328\,d}\over {729}} & {{20}\over 9} -\
  {{100\,u}\over {27}} + {{8\,d}\over {27}} & {{140}\over {27}} +\
  {{908\,u}\over {729}} - {{616\,d}\over {729}} & -{{28}\over 9} +\
  {{148\,u}\over {27}} + {{40\,d}\over {27}} \msvs
\end{array}
\right)
\end{displaymath}

\clearpage
\newsection{Figures of Feynman Diagrams}
%
%
\begin{figure}[h]
\vspace{0.15in}
\centerline{
\epsfysize=1.4in
\epsffile{fig1ew.ps}
}
\vspace{0.15in}
\caption[]{
\label{fig:1}}
\end{figure}

\begin{figure}[h]
\vspace{0.15in}
\centerline{
\epsfysize=1.4in 
\epsffile{fig2ew.ps}
}
\vspace{0.15in}
\caption[]{
\label{fig:2}}
\end{figure}

\begin{figure}[h]
\vspace{0.15in}
\centerline{
\epsfysize=1.4in 
\epsffile{fig3ew.ps}
}
\vspace{0.15in}
\caption[]{
\label{fig:3}}
\end{figure}

\newpage

\begin{figure}[h]
\vspace{0.15in}
\centerline{
\epsfysize=8in
\epsffile{fig4ew.ps}
}
\vspace{0.15in}
\caption[]{
\label{fig:4}}
\end{figure}

\newpage

\begin{figure}[h]
\vspace{0.15in}
\centerline{
\epsfysize=6in
\epsffile{fig5ew.ps}
}
\vspace{0.15in}
\caption[]{
\label{fig:5}}
\end{figure}

\clearpage
\centerline{\Large\bf Figure Captions}

\bigskip

\begin{description}
\item[Figure 1:]
The three basic ways of inserting a given operator into a four--point
function: (a) current--current-, (b) type 1 penguin-, (c) type 2
penguin-insertion.  The wavy lines denote gluons or photons. The 4-vertices
``$\otimes\ \otimes$'' denote standard operator insertions.
\item[Figure 2:]
One--loop current--current diagrams contributing to $\gem$.
The meaning of lines and vertices is the same as in fig.~1.
Possible left-right or up-down reflected diagrams are not shown.
\item[Figure 3:]
One--loop type 1 and 2 penguin diagrams contributing to $\gem$.
The meaning of lines and vertices is the same as in fig.~1.
\item[Figure 4:]
Two--loop current--current diagrams contributing to $\gse$.
The meaning of lines and vertices is the same as in fig.~1.
Possible left-right or up-down reflected diagrams are not shown.
\item[Figure 5:]
Two--loop penguin diagrams contributing to $\gse$.
The wavy lines denote gluons or photons. Square-vertices stand for type 1 and 2
penguin insertions as of figs.~\ref{fig:1}(b) and (c), respectively.
Possible left-right reflected diagrams are not shown.
\end{description}


\newpage
\begin{thebibliography}{10}

\bibitem{gaillard:74}
{\sc M.~K. Gaillard} and {\sc B.~W. Lee},
\newblock {\em Phys. Rev. Lett.} {\bf 33} (1974) 108.

\bibitem{altarelli:74}
{\sc G.~Altarelli} and {\sc L.~Maiani},
\newblock {\em Phys. Lett.} {\bf 52B} (1974) 351.

\bibitem{vainshtein:77}
{\sc A.~I. Vainshtein}, {\sc V.~I. Zakharov}, and {\sc M.~A. Shifman},
\newblock {\em JEPT} {\bf 45} (1977) 670.

\bibitem{gilman:79}
{\sc F.~J. Gilman} and {\sc M.~B. Wise},
\newblock {\em Phys. Rev.} {\bf D20} (1979) 2392.

\bibitem{guberina:80}
{\sc B.~Guberina} and {\sc R.~D. Peccei},
\newblock {\em Nucl. Phys.} {\bf B163} (1980) 289.

\bibitem{bijnenswise:84}
{\sc J.~Bijnens} and {\sc M.~B. Wise},
\newblock {\em Phys. Lett.} {\bf 137 B} (1984) 245.

\bibitem{burasgerard:87}
{\sc A.~J. Buras} and {\sc J.-M. G\'{e}rard},
\newblock {\em Phys. Lett.} {\bf 192B} (1987) 156.

\bibitem{sharpe:87}
{\sc S.~R. Sharpe},
\newblock {\em Phys. Lett.} {\bf 194B} (1987) 551.

\bibitem{lusignoli:89}
{\sc M.~Lusignoli},
\newblock {\em Nucl. Phys.} {\bf B325} (1989) 33.

\bibitem{flynn:89}
{\sc J.~M. Flynn} and {\sc L.~Randall},
\newblock {\em Phys. Lett.} {\bf 224B} (1989) 221,
\newblock Erratum {\bf 235B} (1990) 412.

\bibitem{buchallaetal:90}
{\sc G.~Buchalla}, {\sc A.~J. Buras}, and {\sc M.~K. Harlander},
\newblock {\em Nucl. Phys.} {\bf B337} (1990) 313.

\bibitem{altarelli:81}
{\sc G.~Altarelli}, {\sc G.~Curci}, {\sc G.~Martinelli}, and {\sc S.~Petrarca},
\newblock {\em Nucl. Phys.} {\bf B187} (1981) 461.

\bibitem{burasweisz:90}
{\sc A.~J. Buras} and {\sc P.~H. Weisz},
\newblock {\em Nucl. Phys.} {\bf B333} (1990) 66.

\bibitem{burasetal:92a}
{\sc A.~J. Buras}, {\sc M.~Jamin}, {\sc M.~E. Lautenbacher}, and {\sc P.~H.
  Weisz},
\newblock {\em Nucl. Phys.} {\bf B370} (1992) 69; addendum ibid.~{\em
  Nucl.~Phys.}~{\bf B375} (1992) 501.

\bibitem{burasetal:92b}
{\sc A.~J. Buras}, {\sc M.~Jamin}, and {\sc M.~E. Lautenbacher},
\newblock Two--Loop Anomalous Dimension Matrix for $\dS$ Weak Non-Leptonic
  Decays {\rm I}: ${\cal O}(\as^{2})$,
\newblock {\em Technical University Munich preprint, {\bf TUM-T31-18/92};
  Max-Planck-Institut preprint, {\bf MPI-PAE/PTh 106/92}} .

\bibitem{trueman:79}
{\sc T.~L. Trueman},
\newblock {\em Phys. Lett.} {\bf 88B} (1979) 331.

\bibitem{barrosoetal:92}
{\sc A.~Barroso}, {\sc M.~A. Doncheski}, {\sc H.~Grotch}, {\sc J.~G.
  K{\"o}rner}, and {\sc K.~Schilcher},
\newblock {\em Phys. Lett.} {\bf 261B} (1992) 123.

\bibitem{burasetal:92d}
{\sc A.~J. Buras}, {\sc M.~Jamin}, and {\sc M.~E. Lautenbacher},
\newblock Effective Hamiltonians for $\dS$ and $\dB$ Non-Leptonic Decays beyond
  Leading Logarithms in the Presence of Electroweak Penguins,
\newblock {\em Technical University Munich preprint, in preparation\/, {\bf
  TUM-T31-35/92}} .

\end{thebibliography}
\end{document}